\documentclass[aps,prl,twocolumn,superscriptaddress,showpacs,preprintnumbers,
amsmath,amssymb,showkeys,floatfix,nofootinbib,reprint]{revtex4}

\usepackage{amsmath}
\usepackage{amsfonts}
\usepackage{amssymb}
\usepackage{graphicx}
\usepackage{color}
\usepackage{bm}
\usepackage{hyperref}

\begin{document}

%
%
%

\def\ada#1{\textcolor{blue}{#1}}
\def\jonas#1{\textcolor{red}{#1}}

\def\ket#1{ $ \left\vert  #1   \right\rangle $}
\def\ketm#1{  \left\vert  #1   \right\rangle   }
\def\bra#1{ $ \left\langle  #1   \right\vert $ }
\def\bram#1{  \left\langle  #1   \right\vert   }
\def\spr#1#2{ $ \left\langle #1 \left\vert \right. #2 \right\rangle $ }
\def\sprm#1#2{  \left\langle #1 \left\vert \right. #2 \right\rangle   }
\def\me#1#2#3{ $ \left\langle #1 \left\vert  #2 \right\vert #3 \right\rangle $}
\def\mem#1#2#3{  \left\langle #1 \left\vert  #2 \right\vert #3 \right\rangle   }
\def\redme#1#2#3{ $ \left\langle #1 \left\Vert
                  #2 \right\Vert #3 \right\rangle $ }
\def\redmem#1#2#3{  \left\langle #1 \left\Vert
                  #2 \right\Vert #3 \right\rangle   }
\def\threej#1#2#3#4#5#6{ $ \left( \matrix{ #1 & #2 & #3  \cr
                                           #4 & #5 & #6  } \right) $ }
\def\threejm#1#2#3#4#5#6{  \left( \matrix{ #1 & #2 & #3  \cr
                                           #4 & #5 & #6  } \right)   }
\def\sixj#1#2#3#4#5#6{ $ \left\{ \matrix{ #1 & #2 & #3  \cr
                                          #4 & #5 & #6  } \right\} $ }
\def\sixjm#1#2#3#4#5#6{  \left\{ \matrix{ #1 & #2 & #3  \cr
                                          #4 & #5 & #6  } \right\} }

\def\ninejm#1#2#3#4#5#6#7#8#9{  \left\{ \matrix{ #1 & #2 & #3  \cr
                                                 #4 & #5 & #6  \cr
                         #7 & #8 & #9  } \right\}   }
%
%
%
%

%
%

\title{Dominant Secondary Nuclear Photoexcitation with the X-ray Free Electron Laser}

%
%

\author{Jonas \surname{Gunst}}
\email{Jonas.Gunst@mpi-hd.mpg.de}
\affiliation{Max-Planck-Institut f\"ur Kernphysik, Saupfercheckweg 1, D-69117 Heidelberg, Germany}

\author{Yuri A. \surname{Litvinov}}
\affiliation{GSI Helmholtzzentrum f{\"u}r Schwerionenforschung GmbH, 64291 Darmstadt, Germany}

\author{Christoph H. \surname{Keitel}}
\affiliation{Max-Planck-Institut f\"ur Kernphysik, Saupfercheckweg 1, D-69117 Heidelberg, Germany}

\author{Adriana \surname{P\'alffy}}
\email{Palffy@mpi-hd.mpg.de}
\affiliation{Max-Planck-Institut f\"ur Kernphysik, Saupfercheckweg 1, D-69117 Heidelberg, Germany}


\date{\today}

%
%
%
%
%
%
%
\begin{abstract}

The new regime of resonant nuclear photoexcitation rendered possible by  x-ray free electron laser beams interacting with solid state targets is investigated theoretically. Our results unexpectedly show that  secondary processes coupling nuclei to the atomic shell in the created cold high-density plasma  can dominate direct photoexcitation. As an example we discuss the case of $^{93\mathrm{m}}$Mo isomer depletion for which nuclear excitation by electron capture as secondary process is shown to be orders of magnitude more efficient than the direct laser-nucleus interaction. General arguments revisiting the role of the x-ray free electron laser in nuclear experiments involving solid-state targets are further deduced.

\end{abstract}
%
%






\maketitle


The new X-ray Free Electron Laser (XFEL) facilities \cite{LCLS-web,SACLA-web} may provide both the x-ray photon energies  and the very high brilliance required for resonant driving of nuclear transitions \cite{Baldwin1997.R,Schwoerer2006.Book,Burvenich2006.PRL,Palffy2008.PRC,Rabitz2011.PRA,Olga2011.PRA,DiPiazza2012.RMP,Adams2013.JoMO}.  So far, the resonant interaction between nuclei and the electromagnetic field was studied  in experiments performed with broadband synchrotron radiation (SR) \cite{Gerdau1999.HI} or bremsstrahlung \cite{Kneissl1996}. 
The peak brilliance of XFEL light reaches up to eight orders of magnitude higher than that of SR sources \cite{Tesla-TDR-2001} and is expected to bring significant progress in light-nucleus interaction experiments.  In particular SR experiments with 
M\"{o}ssbauer solid-state targets, mostly involving $^{57}$Fe  \cite{Shvydko1996.PRL,Rohlsberger2010.S,Rohlsberger2012.N}, provide only weak nuclear excitation despite the high target density and could benefit from the XFEL intensity. While so far in these experiments the electronic response only acted as background, the increase of the electric field strength leads to drastic changes in the interaction between photons and electrons which may additionally influence the nuclear excitation. Due to the unique interaction between high intensity x--ray pulses and matter \cite{Young2010.N,Hau-Riege.book} new states like cold, high--density plasmas can originate \cite{Lee2003.JOSAB,Vinko2012.N}. In such environments secondary nuclear processes from the coupling  to the atomic shell are rendered possible by the presence of free electrons and atomic vacancies. This is also a new and diametrically opposed situation compared to photonuclear studies involving petawatt optical lasers 
\cite{KenL2000.PRL,Cowan2000.PRL,Gibbon2005.Book,Spohr2008.NJP,Mourou2011.S}.

In this Letter we investigate the nuclear excitation induced by the XFEL pulse 
shining on a nuclear solid-state target. We show that surprisingly, secondary nuclear excitation by electron capture (NEEC) in the occurring plasma  can exceed by orders of magnitude the direct resonant photoexcitation despite the laser photons being tuned on the nuclear transition. Furthermore,  we find that NEEC is more robust since it is less sensitive to the laser photon frequency fulfilling the resonance condition. This is a new feature as  electronic processes were not relevant for experiments performed with SR, where the fast electronic response of the sample was negotiated by time gating \cite{Roehlsberger2004,Sturhahn2004.J}.  The concrete example studied here is the case of photoexcitation starting from the 6.85~h long-lived isomeric state of $^{93}$Mo at approx.~2.5 MeV excitation energy. The energy stored in the isomer may be released on demand by driving a 4.85~keV $E2$ transition in an isomer triggering scenario \cite{Walker1999.N, Aprahamian2005.NP,Belic1999.PRL,Collins1999.PRL,Belic2002.PRC,
Carroll2004.L,Palffy2007.PRL}. Due to the advantageous energy ratio
and available x-ray light sources, isomer depletion in $^{93}$Mo opens 
interesting prospects for the development of nuclear energy storage solutions. We show that for present XFEL parameters, the secondary nuclear excitation via NEEC is dominant and investigate under which conditions is this a general feature of the XFEL interaction with solid-state nuclear targets.

\begin{figure}
\centering
\includegraphics[width=1.0\linewidth]{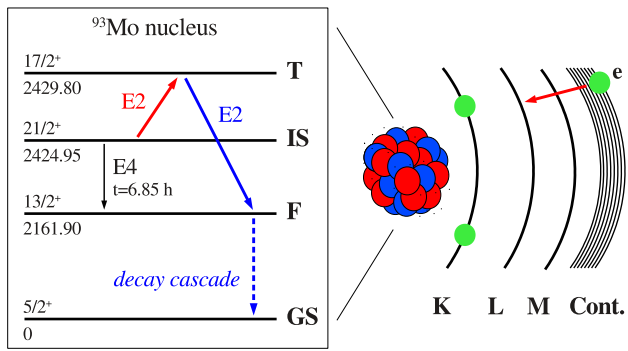}%
\caption{(Color online) $^{93\mathrm{m}}$Mo excitation induced by NEEC into the $L$ shell (right) with subsequent decay to the nuclear ground state (long blue solid and dashed arrows in the left panel). The nuclear levels are labeled with their total angular momentum, parity  and energy (in keV).
\label{fig:neec}}
\end{figure}

In the resonant process of NEEC,  a free electron is captured into a bound atomic  state by the simultaneous excitation of the nucleus \cite{Palffy2010.CP} as illustrated in Fig.~\ref{fig:neec}  for  ${}^{93\mathrm{m}}$Mo.  The nucleus is initially in its ground state or an excited metastable state, i.e., an isomer.  The idea of isomer triggering is to connect the long--lived isomeric state with a higher level linked to freely radiating states as shown in Fig.~\ref{fig:neec}, in order to release the stored nuclear excitation energy on demand. In the case of $^{93}$Mo, a 4.85~keV $E2$ triggering transition exists \cite{ENSDF,Hasegawa2011.PLB} making this isomer particularly attractive for an XFEL--induced activation. Moreover, in 
$^{93}_{41}$Nb(p,n)$^{93\mathrm{m}}_{\phantom{m} 42}$Mo reactions \cite{exfor},  the  isomers can be produced  directly embedded into 1~$\mu$m thick solid--state niobium foils \cite{supplement}, providing high-density targets.   Most of the XFEL photons will however interact with the atomic shells producing ionization and leading to plasma generation \cite{Hau-Riege.book}. The study of NEEC in plasmas was so far restricted to astrophysical environments \cite{Gosselin2004.PRC,Morel2004.PRA,Gosselin2007.PRC,Morel2010.PRC} or optical-laser-generated plasmas \cite{Harston1999.PRC} where no equivalent of the direct photoexcitation channel under investigation here exists.

Currently, there are two operating XFEL facilities worldwide, the Linac Coherent Light Source (LCLS) at SLAC in Stanford, USA, \cite{LCLS-web} and the SPring-8 Angstrom Compact free electron laser (SACLA) in Japan \cite{SACLA-web}, both capable to provide 4.85 keV photons. The parameters of interest for our case are given in Table~\ref{table:photoexcitation}. In addition, we also consider parameters for the European XFEL under construction at DESY in Hamburg, Germany \cite{EurXFEL-tdr}. The light--nucleus coupling can be  described theoretically via the density matrix formalism using the semi-classical approach  \cite{ScullyZubairy}. Typically, the nuclear decay happens on much slower time scales than the XFEL pulse duration,  enabling us to restrict the interaction model to the two upper states in Fig.~\ref{fig:neec}. As decoherent relaxation processes we include all spontaneous decay channels governed by radiative decay and internal conversion (IC), as well as a field-dephasing term that accounts for the 
limited coherence time. The large discrepancy between the laser width $\Gamma_{\mathrm{xfel}}$ ($\sim$10~eV) and the nuclear transition width $\Gamma_{n}$ ($\sim$10$^{-7}$~eV) allows only a small fraction of the laser photons to be resonant with the nuclear transition and justifies the introduction of effective laser parameters \cite{Palffy2008.PRC}.

\renewcommand{\arraystretch}{1.2}
\begin{table}
  \centering
  \begin{tabular}{lccc}
  \hline\hline
  & & &  \tabularnewline[-0.4cm]
  Parameter & LCLS  \cite{LCLS-tdr,Emma2010.NP,Gutt2012.PRL}& SACLA \cite{SACLA-tdr,Ishikawa2012.NP}& Eur.~XFEL \cite{EurXFEL-tdr} \tabularnewline
  & & & \tabularnewline[-0.4cm] \hline
  & & & \tabularnewline[-0.4cm]
  $E_{\mathrm{max}}$ (eV) & 10332 & 19556 & 24800 \tabularnewline
  $BW$ & 2$\times$10$^{-3}$ & 2.2$\times$10$^{-3}$ & 8$\times$10$^{-4}$ \tabularnewline
  $T_{\rm coh}$ (fs) & 2 & --\footnote{In our calculations we assumed 10$\%$ of the pulse duration, i.e., 10~fs.} & 0.2 \tabularnewline
  $I$ (W/cm$^2$) & 3.9$\times$10$^{17}$ & 9.8$\times$10$^{16}$ & 2.0$\times$10$^{17}$ \tabularnewline
  $f_{\rm rep}$ (Hz) & 30 & 10 & 4$\times$10$^{4}$ \tabularnewline 
  & & & \tabularnewline[-0.4cm] \hline
  & & & \tabularnewline[-0.4cm] 
  $\rho_{\mathrm{trig}}$ & 1.8$\times$10$^{-20}$ & 1.7$\times$10$^{-20}$ & 2.4$\times$10$^{-21}$ \tabularnewline
  & & & \tabularnewline[-0.4cm] \hline\hline
  \end{tabular}
  \caption{The maximal achievable photon energy $E_{\rm max}$, bandwidth $BW$, coherence time $T_{\rm coh}$, peak intensity $I_{}$, pulse repetition rate $f_{\rm rep}$, and calculated triggering occupation number $\rho_{\mathrm{trig}}$ after the photoexcitation of a single XFEL pulse for the three considered XFEL facilities.}
  \label{table:photoexcitation}
\end{table}
\renewcommand{\arraystretch}{1}

The dynamics of the nuclear occupation numbers is determined by the  Bloch equations which can be written in terms of the interaction matrix elements \me{T}{H_I}{IS}, where $H_I$ represents the light--nucleus interaction, \ket{IS} the isomeric state and \ket{T} the triggering level. This in turn can be related in the long-wavelength-approximation \cite{RingSchuck} to the reduced nuclear transition probabilities \cite{Hasegawa2011.PLB,ENSDF}. Following the formalism presented in Ref.~\cite{Palffy2008.PRC}, the Bloch equations are solved numerically. Results for the excited state occupation number $\rho_{\mathrm{trig}}$, i.e., the fraction of excitation produced per laser pulse per nucleus in the sample,  are shown in Table \ref{table:photoexcitation}.  The calculations were performed considering a pulse duration of 100~fs and a focal spot of 10~$\mu$m$^2$. The highest value $\rho_{\mathrm{trig}}=1.8\times$10$^{-20}$ is achieved for the LCLS parameters.

A common problem of all presently operating XFELs is the poor temporal coherence (indicated by the coherence time $T_{\rm coh}$ in Table \ref{table:photoexcitation}) due to random fluctuations in the initial electron charge density. The corresponding decoherence rate in the Bloch equations limits substantially the possible magnitude of $\rho_{\mathrm{trig}}$. At the moment, there are two proposals how to raise the temporal coherence of XFELs: (i) load the undulator with an already seeded light pulse in order to reduce shot--to--shot fluctuations at the start--up (seeded XFEL) \cite{Feldhaus1997.OC,Saldin2001.NIaMiPRA,DESY-seeding}; (ii) construct an x--ray cavity based on diamond mirrors \cite{Shvydko2010.NP,Shvydko2011.NP} which directs the light pulse several times through the undulator (XFEL oscillator - XFELO) \cite{Kim2008.PRL}. Totally coherent x--ray pulses increase the triggering occupation number by at least 4 up to 6 orders of magnitude in comparison to their unseeded counterparts presented in 
Table \ref{table:photoexcitation}. Calculations with expected functioning parameters for the XFELO (with a pulse duration of 1~ps)  deliver a value as high as  4.4$\times$10$^{-14}$ for $\rho_{\rm trig}$, proving the importance of the temporal coherence.

The nuclear excitation induced directly by the laser should be compared to its secondary electronic-processes-induced counterpart. 
The  interaction of XFEL light with the electrons of the metallic target causes  the direct production of inner shell holes, the uniform radiation of the sample and a rapid heating process \cite{Hau-Riege.book}, eventually leading to the formation of a plasma with unique properties, like uniform electron temperature and almost solid-state density \cite{Lee2003.JOSAB}. In this environment, NEEC takes place on a longer time scale compared to the laser pulse duration, as long as suitable free electrons and atomic vacancies are available. In our estimate we consider this to hold for 100 ps after the laser pulse as long as the mass transport of the inner core plasma can  be yet neglected \cite{Hau-Riege.book}. 

The plasma evolution is modeled in two stages. First, plasma generation is dominated by the photoionization of inner shell electrons and the subsequent refilling of the arising holes by either radiative or Auger decays \cite{Vinko2012.N}. Note that in our scenario only holes in the  $L$ shell and above can be produced by the laser photons due to the high $K$--shell ionization potentials. We estimate the laser energy deposited into the sample with the help of mass photoabsorption coefficients \cite{Henke1993.ADaNDT} which are in the first approximation held constant. By further accounting for energy conservation of the inner shell photoionization and the first sequence of Auger decays, the 
averaged electron temperature can be estimated \cite{Tesla-TDR-2001}.  As the next step following the XFEL-induced plasma formation, radiative and collisional processes begin to dictate the further dynamics, and fast thermalization occurs. These are accounted for with the help of the population kinetics model implemented in the FLYCHK code \cite{FLYCHK}. The latter is based on a collisional radiative model and requires as input the averaged electron temperature $T_e$ and the ion density. In our case we estimated $T_e=350$~eV for the 1~$\mu$m thick Mo/Nb--target and $10^{12}$ resonant laser photons with 10~$\mu$m$^2$ focus, and assumed the Nb solid-state density value. The electron density $n_e$ which adds up to 1.3$\times$10$^{24}$~cm$^{-3}$ and the charge state distribution (CSD) in the plasma are obtained using the FLYCHK code. 

The total net NEEC rate in the plasma relies on the available charge states and electron energies. Performing a summation over the available charge states leads to
\begin{equation}
\lambda_{\rm neec} = \sum_q P_q \lambda^{q}_{\rm neec},
\label{eq:neec.total}
\end{equation}
where $P_q$ denotes the probability to find an ion in the plasma with initial charge state $q$ before the capture. The partial NEEC rate $\lambda_{\rm neec}^q$ contains a second summation over all contributing electronic capture levels \ket{\alpha_d} and considers the available electron distribution $\phi_e(E)$ by a convolution integral over the  electron kinetic energy $E$,
\begin{equation}
\lambda^{q}_{\rm neec} = \sum_{\alpha_d} P_{\alpha_d} \int dE \, \sigma_{\rm neec}^{i \rightarrow d}(E) \phi_e(E).
\label{eq:neec.partial}
\end{equation}
Here, \ket{i} represents the initial state composed of the isomeric level \ket{IS} and the initial electronic state, and \ket{d} the intermediate level determined by \ket{T} and \ket{\alpha_d}. The probability to find \ket{\alpha_d} unoccupied is denoted by $P_{\alpha_d}$. For a given charge state $q$ of the ion, we assume capture into free orbitals of the ion ground state only, i.e., in our case  $P_{\alpha_d}=1$.
Furthermore, in the derivation of Eq.~(\ref{eq:neec.partial}) the single resonance approximation was applied, which allows to write the total NEEC cross section in terms of the single resonances $\sigma_{neec}^{i \rightarrow d}$. The dependence of the latter on  the free electron energy is given by \cite{Palffy2006.PRA} (in a.u.)
\begin{equation}
\sigma_{\rm neec}^{i \rightarrow d} (E)= \frac{2\pi^2}{p^2} Y_{\rm neec}^{i \rightarrow d}L_d (E-E_d)\, ,
\label{eq:neec.cs}
\end{equation}
where $p$ is the momentum of the recombining electron, $Y_{\rm neec}^{i \rightarrow d}$ the NEEC transition width and $L_d (E-E_{d})$ the Lorentzian profile centered around the resonance energy $E_{d}$ with a width given by the natural width $\Gamma_d$ of the level \ket{d}. The momentum dependence implies that NEEC favors the capture into deep vacancies. The natural width of the resonance is given by the nuclear transition width if capture occurs into the electronic ground state, whereas otherwise also the width of the electronic bound initial state must be taken into account. The calculation of the NEEC width \cite{Palffy2006.PRA} involves the nuclear reduced transition probabilities and electronic matrix elements which we evaluate considering the single active electron approximation. The required many--electron wave functions, energies and transition widths are obtained using the relativistic Multi-Configurational Dirac-Fock method implemented in GRASP92 \cite{GRASP}.

\begin{figure}
\centering
\includegraphics[width=1.0\linewidth]{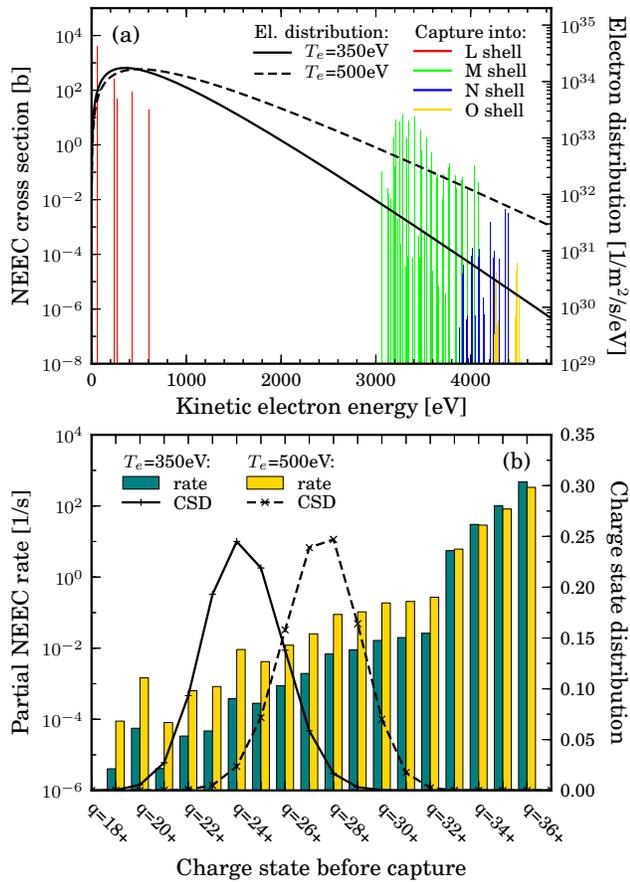}%
\caption{(Color online) (a) NEEC resonance cross sections $\sigma_{\rm neec}$ for captures into the $L$, $M$, $N$ and $O$ shell (left axis) together with the electronic energy distribution (right axis). (b) Partial NEEC rate $\lambda^{q}_{\rm neec}$ (left axis) together with the corresponding CSD (right axis). Results are presented for two plasma temperatures, 350 eV and 500 eV.
\label{fig:neec-results}}
\end{figure}

Results for the case of ${}^{93\mathrm{m}}$Mo triggering are shown in Fig.~\ref{fig:neec-results}(a), where the single-resonance cross sections for several electron configurations undergoing NEEC into the $L$, $M$, $N$ and $O$ shells and the electron distributions for $T_e$=350~eV and $T_e$=500~eV are plotted as a function  of the  electron kinetic energy. Both the cross section and the number of available resonant electrons decrease with increasing $E$ or correspondingly by going to higher capture channels. Therefore, a cut-off level can be found for each charge state $q$ starting with which the NEEC excitation can be neglected. This procedure reduces the contributing resonances in Eq.~(\ref{eq:neec.partial}) to a finite set of atomic levels, and  provides a lower limit for $\lambda_{\rm neec}$.

In Fig.~\ref{fig:neec-results}(b) the partial NEEC rates $\lambda^{q}_{\rm neec}$ and the CSD are presented for $T_e$=350~eV and for comparison also for $T_e$=500~eV for charge states from $q$=19+ to $q$=36+. The excitation probability increases with $q$, since deeper vacancies are present in higher charge states. Moreover, the  CSD shows that only a limited number of charge states is available for a given plasma temperature $T_e$. For instance in the case of $T_e$=350~eV, NEEC into the $L$ shell is forbidden since this would require a Mo charge state of at least $q$=33+. Higher $q$-values can be reached by raising the plasma temperature, which can  practically be implemented  by turning the laser off resonance to higher frequencies \cite{Vinko2012.N}. With higher  charge states available and an increase in the overall number of resonant electrons, the total NEEC rate $\lambda_{\rm neec}$ increases with the plasma temperature $T_e$.

For $T_e$=350~eV the NEEC excitation entails a triggering occupation number of $\rho_{\mathrm{trig}}=5.5\times$10$^{-14}$ per pulse which is about 6 orders of magnitude higher than the earlier indicated photoexcitation values for available XFEL parameters and is even competitive with the XFELO results. Already the comparison between the theoretical cross sections of NEEC and photoexcitation \cite{Palffy2007.PRL,Palffy2008.JoMO} indicated that the coupling to the atomic shell can be more effective than to the radiation field for low--lying triggering levels. However, the main reason for the deviation between these two excitation mechanisms in our scenario is the discrepancy between the interaction times. While the photon-nucleus coupling is confined to the XFEL pulse duration ($\sim$100~fs), the NEEC process takes place during the whole plasma lifetime which can be orders of magnitude longer. The total NEEC rates are also affected by other plasma parameters. Our estimates show a strong dependence of $\lambda_{
\rm neec}$ on 
the temperature $T_e$ and a weaker sensitivity towards the electron density $n_e$. For instance, a reduction of 10\% in $T_e$ leads to a four times smaller NEEC rate, whereas a tenfold increase of the plasma volume results in a modification in $\lambda_{\rm neec}$ of only 8\%. 
More extended, non-equilibrium calculations of the plasma parameters, especially for the electron temperature, could even open the way for using the NEEC excitation rate as a plasma diagnosis tool in future experiments.

Clearly measurable signatures for the nuclear excitation process are desirable for experiments. In the example of ${}^{93\rm m}$Mo triggering,  an outstandingly high energetic photon of 1~MeV is emitted in the decay cascade from the triggering level to the ground state regardless of the activation mechanism. Due to its high energy, such a $\gamma$-ray photon is unlikely to have electronic origin and can leave the plasma environment undisturbed. Although the emission may occur during the hydrodynamic expansion or even Coulomb explosion of the sample (lifetime of \ket{T} is about 3~ns) the preserved photonic energy of 1~MeV can serve as distinct signal for the triggering process.

Due to its resonant nature, the direct photoexcitation channel can be switched on and off by tuning the laser in- or out-of-resonance, respectively. However, in the case of ${}^{93}$Mo, the 4.85~keV triggering transition energy has at present an uncertainty of 80~eV \cite{ENSDF}, which is a limiting factor for the experimental realization of  direct photoexcitation given the eV XFEL bandwidth. In contrast, NEEC is not sensitive to the resonance condition because it takes place in a plasma environment where a broad electron energy distribution opens many resonance channels. Thus, not only is the NEEC triggering mechanism dominating over the direct channel, but it is also significantly more robust.

Considering the $^{93}_{41}$Nb(p,n)$^{93\mathrm{m}}_{\phantom{m} 42}$Mo reaction cross section \cite{exfor}, a ${}^{93\rm m}$Mo isomer density of 10$^{16}$~cm$^{-3}$  can be achieved in the solid--state Nb foils \cite{supplement} using standard proton 
beams like the LINAC at GSI \cite{GSI-web,LINAC-web}. For the resonant driving assuming the European XFEL parameters we obtain 9.6$\times$10$^{-12}$ and 2.2$\times$10$^{-4}$ signal photons/s induced by direct photoexcitation and NEEC, respectively, for a single $\mu$m Nb target foil. Signal rates calculated for LCLS parameters are more than two orders of magnitude smaller, namely 5.6$\times$10$^{-14}$ and 1.7$\times$10$^{-7}$ signal photons/s, mainly due to the lower repetition rate. The XFEL  pulse repetition can be exploited by using a tape-station system target. Moreover, it is possible to gain  at least one order of magnitude in signal  by using a stack of target foils as long as the laser intensity attenuation does not prohibit the plasma formation. Nevertheless, it would be desirable to increase the isomeric density in the laser focal spot, for instance by increasing the intensity or by a stronger focusing of the proton beam.

The generalization of our results for the excitation of other low-lying nuclear states (a suitable list is provided in Ref.~\cite{Junker2012.NJoP}) via XFEL light relies on three aspects. Firstly, the interaction time available for the plasma electrons to excite the nucleus depends on the hydrodynamic expansion time scale and can be orders of magnitude longer than the pulse duration over which resonant photons are present. Secondly, for the small excitation energies that can be reached with available and forthcoming XFEL machines ($E\le$ 25~keV), the NEEC cross sections are larger than the photoexcitation ones, as shown also by the large values ($\gg 1$) of the IC coefficients \cite{ENSDF}, i.e., the ratio between 
the rates of the corresponding inverse processes IC and $\gamma$ decay.  Finally, the XFEL photons resonant with the nuclear transition  create in large numbers charge states and free electrons that are not far from the resonant NEEC condition. Only in the limit of high nuclear transition energies like for ${}^{57}$Fe (14.4~keV), ${}^{149}$Sm (22.5~keV) or ${}^{119}$Sn (23.9~keV), is the atomic photoabsorption cross section so strongly reduced that the NEEC resonance energies for the available charge states are expected rather at the falling high-energy tail of the electron distribution resulting in a low NEEC rate $\lambda_{\rm neec}$. In all other cases listed in Ref.~\cite{Junker2012.NJoP}, inner shell vacancies characterized by high NEEC recombination cross sections will be produced which renders the successful capture of the generated low-energy plasma electrons possible. In addition,  secondary nuclear excitation in the plasma is all the more to be considered in view of 
the typically high uncertainties of the nuclear transition energies  which reduce from the start the chances for direct resonant photoexcitation. 

We would like to thank D. Bauer for fruitful discussions. This research was in part supported by the Helmholtz--CAS Joint Research Group HCJRG--108.

\bibliographystyle{apsrev-no-url-issn}
\bibliography{mybibJ}{}

\end{document}